# Quantum and Classical Information Theory with Disentropy

Rubens Viana Ramos

rubens.ramos@ufc.br

*Lab. of Quantum Information Technology, Department of Teleinformatic Engineering – Federal University of Ceara - DETI/UFC, C.P. 6007 – Campus do Pici - 60455-970 Fortaleza-Ce, Brazil.*

*Abstract*

Entropy is a famous and well established concept in physics and engineering that can be used for explanation of basic fundamentals as well it finds applications in several areas, from quantum physics to astronomy, from network communication to medical image processing, for example. Now, entropy meets its dual, the disentropy. As such, the disentropy can be used everywhere entropy is used, offering a different point of view: since entropy is a measure of disorder or uncertainty, disentropy is a measure of order or certainty. Thus, important concepts of physics can be rewritten using disentropy instead of entropy. Although there is a large range of problems that can be solved using entropy or disentropy, there are situations where only the disentropy can be used. This happens because the disentropy can provide a real output value when its argument is negative, while the entropy cannot. Thus, it is possible to calculate, for example, the disentropy of quasi-probability distributions like the Wigner function of highly quantum states. In this direction, the present work shows applications of the disentropy in a small list of problems: quantum and classical information theory, black hole thermodynamics, image processing and number theory.

*Key words* – Lambert-Tsallis $W_q$ function; disentropy; quantum and classical information theory; black hole, image processing; integer numbers

## 1. Introduction

Entropy is a famous and well established concept in physics and engineering that can be used for the explanation of basic fundamentals as well it finds applications in several areas like classical and quantum information theory, plasma physics, turbulence, astrophysics, cosmology and black holes, nonlinear dynamical systems, glasses and spin-glasses, astronomy, solar physics, image processing, cryptography, complex networks, among others [1 and references there in]. It is well established that the entropy of a system can be understood as a measure of disorder or uncertainty of that system. If a system is not maximally disordered, this means that it has some order.

Hence, it is also natural the existence of a measure of order or certainty. This measure is named *disentropy*, since it plays the opposite role of entropy. Thus, the same system has a degree of disorder measured by entropy and a degree of order measured by disentropy. Due to the second law of thermodynamics, the result of the interplay between entropy and disentropy has to be another entropy: $S_{net} = S_{int} - D$. In other words, the net entropy is equal to the difference between intrinsic entropy and disentropy. It is the net entropy that must obey the second law of thermodynamics. In this work, the focus is the disentropy and its applications. Examples in quantum and classical information theory, black hole thermodynamics, image processing and number theory are considered. However, before to start the discussion about disentropy, it is necessary to provide a brief review of the Lambert functions and its generalizations.

The Lambert $W$ function is an elementary function that has found many applications in different areas of science [2-7]. Basically, it is the solution of the equation

$$W(z)e^{W(z)} = z. \tag{1}$$

Taking the logarithm in both sides of eq. (1), one finds a recursive relation between Lambert and logarithmic functions: $\log(z) = W(z) + \log[W(z)]$. The Lambert $W$ function has infinite branches but only two of them, named $W_0(z)$ and $W_{-1}(z)$, provide real values for real arguments. The branch point is ($z_b = -1/e$, $W(z_b) = -1$). The function $W_0(z)$ is valid in the interval $-1/e \leq z \leq \infty$ while the function $W_{-1}(z)$ is valid in the interval $-1/e \leq z < 0$. The Lambert function has been used to provide analytical solutions for problems where traditionally numerical methods are used. For example, it was shown in [3] that Wien's displacement law is $\lambda_{max}T = (hc/k_B)/(5+W_0(-5e^{-5}))$, where $\lambda_{max}$ is the wavelength at which the spectral distribution of a black body radiation is maximal and $T$ is the absolute temperature ($h$, $k_B$ and $c$ are, respectively, the Planck and Boltzmann constants and the light velocity). On the other hand, in [4] the eigenenergies of a square quantum finite well were found using $W$ while in [7] the $W$ function was used to model nonlinear electronic circuits.

One may note that eq. (1) can be generalized as

$$R_\lambda(z)\lambda^{R_\lambda(z)} = z. \tag{2}$$

Taking the logarithm (basis $\lambda$) in both sides of (2) one gets the recursive relation

$$\log_\lambda(z) = R_\lambda(z) + \log_\lambda[R_\lambda(z)]. \tag{3}$$

For example, for $\lambda = 2$ one has $R_2(z)2^{R_2(z)} = z$ and

$$\log_2(z) = R_2(z) + \log_2[R_2(z)]. \tag{4}$$

One should note that $R_2(z)$ has also two real branches $R_2^0$ and $R_2^{-1}$. Its branch point is $(z_b = -1/\log(2^e), R_2(z_b) = -1/\log(2))$. Figure 1 shows the curves of $W(z)$ and $R_2(z)$.

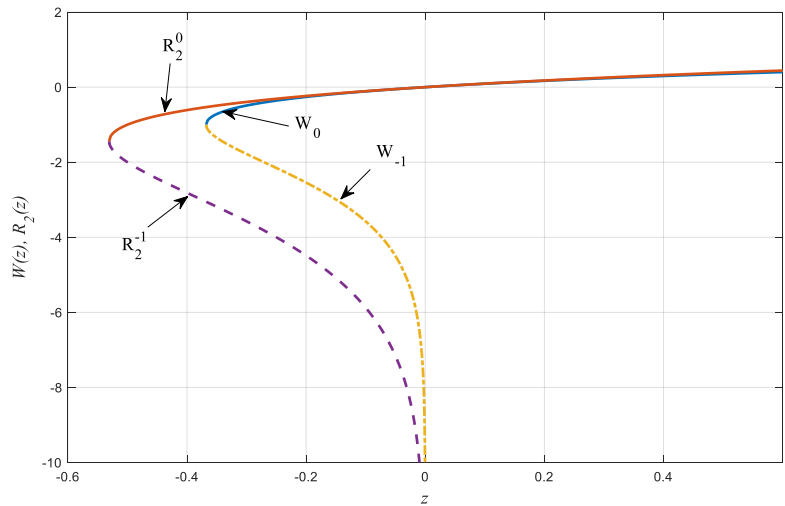

Fig. 1 – $W(z)$ and $R_2(z)$ versus $z$.

Basically, working with (2) one gets

$$R_\lambda(z) = \log_\lambda(e) W\left(\frac{z}{\log_\lambda(e)}\right). \tag{5}$$

Another generalization of $W(z)$ is obtained if the exponential in eq. (1) is changed by the generalized Tsallis $q$-exponential: $\exp_q(z) = [1+(1-q)z]^{1/(1-q)}$ for $q \neq 1$ and $[1+(1-q)z] \geq 0$ [8,9]. Thus, one has the Lambert-Tsallis equation,

$$W_q(z)e_q^{W_q(z)} = z, \tag{6}$$

whose is solution is the Lambert-Tsallis $W_q$ function introduced in [10]. For $q = 1$ one has $\exp_{q=1}(z) = e^z$ and $W_1(z) = W(z)$. The branch point of $W_q(z)$ is ($z_b = \exp_q(1/(q-2))/(q-2)$, $W_q(z_b) = 1/(q-2)$). One can find the analytical expressions for $W_q(z)$ solving the equation [10]

$$x(r+x)_+^r = r^r z, \tag{7}$$

where $r = 1/(1-q)$, $x = W_{(r-1)/r}(z)$ and $(a)_+ = \max\{a,0\}$. The solutions of (7) are roots of polynomials. For example, for $q$ equal to 1/2, 3/2, 4/3 and 2 one has the following upper branches (the upperscript '0' is dropped out) for the respective Lambert-Tsallis functions [11]

$$W_{1/2}^0(z) = \frac{\left[3\sqrt[3]{2z + \sqrt{\left(2z + \frac{8}{27}\right)^2 - \frac{64}{729}} + \frac{8}{27}} - 2\right]^2}{9\sqrt[3]{2z + \sqrt{\left(2z + \frac{8}{27}\right)^2 - \frac{64}{729}} + \frac{8}{27}}}; \quad z \in (-0.29629, \infty) \tag{8}$$

$$W_{3/2}^0(z) = \frac{2(z+1) - 2\sqrt{2z+1}}{z}; \quad z \in (-1/2, \infty) \tag{9}$$

$$W_{4/3}^0(z) = \sqrt[3]{\frac{1}{2}}\sqrt[3]{\sqrt{\frac{6561z + 2916}{z^3}} - \frac{81}{z}} - \frac{9\sqrt[3]{2}}{z\sqrt[3]{\sqrt{\frac{6561z + 2916}{z^3}} - \frac{81}{z}}} + 3; \quad z \in (0, \infty) \tag{10}$$

$$W_2(z) = \frac{z}{z+1}; \quad z \in (-1, \infty). \tag{11}$$

As one may note, $W_2(z)$ does not have a finite branch point.

It is also possible to generalize the Lambert equation by using the Kaniadakis exponential [12]

$$W_\kappa(z) e_\kappa^{W_\kappa(z)} = z, \tag{12}$$

where [13]

$$e_\kappa^z = \left[\sqrt{1+\kappa^2 z^2} + \kappa z\right]^{\frac{1}{\kappa}}. \qquad (13)$$

Introducing $r = 1/\kappa$ and $x = W_{\frac{1}{r}}(z)$ in (12)-(13) one obtains

$$x\left[\sqrt{r^2 + x^2} + x\right]^r = r^r z, \qquad (14)$$

whose solutions (roots) are the Lambert-Kaniadakis functions. The branch point of a Lambert-Kaniadakis function, for $0 \leq \kappa^2 < 1$, is ($z_b = -(1-\kappa)^{(1-\kappa)/2\kappa}/(1+\kappa)^{(1+\kappa)/2\kappa}$, $W_\kappa(z_b) = -(1-\kappa^2)^{-1/2}$). The curves of $W_{q=3/2}(z)$ and $W_{\kappa=1/3}(z)$, for example, can be seen in Fig. 2.

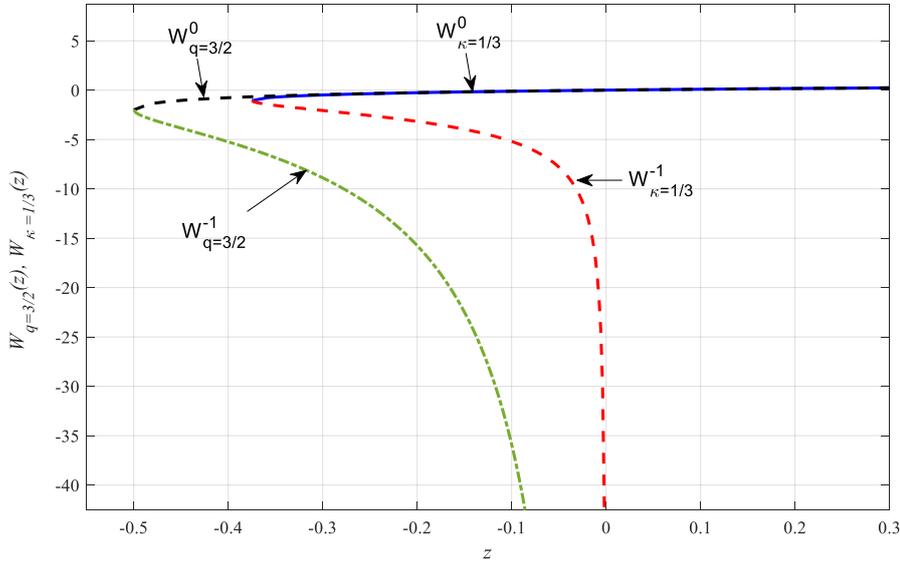

Fig. 2 – $W_{q=3/2}(z)$ and $W_{\kappa=1/3}(z)$ versus $z$.

The most important concept discussed in this work is the disentropy. The mathematical definition of the disentropy is based on the Lambert function and its generalizations. The disentropy related to Tsallis entropy uses the Lambert-Tsallis $W_q$ function [10]. Taking the $q$-logarithm in both sides of eq. (6) one gets

$$\ln_q(x) = W_q(x) +_q \ln_q\left[W_q(x)\right] = W_q(x) + \ln_q\left[W_q(x)\right] + (1-q)W_q(x)\ln_q\left[W_q(x)\right], \quad (15)$$

where '$+_q$' is the $q$-addition operation and $\ln_q(x)$ is the $q$-logarithm function [8,9]. Using (15), the Tsallis entropy of a discrete variable can be rewritten as

$$S_q = -\sum_n p_n^q \ln_q(p_n) = -\sum_n p_n^q W_q(p_n) - \sum_n p_n^q \ln_q(W_q(p_n)) - (1-q)\sum_n p_n^q W_q(p_n) \ln_q(W_q(p_n)). \quad (16)$$

The term

$$D_q = \sum_n p_n^q W_q(p_n) \quad (17)$$

in (16) is positive and it is the mathematical definition of the disentropy since it is maximal for a delta distribution and minimal for a uniform distribution [10], hence it is a measure of order or certainty. When $q = 1$ eq. (16) can be written as $S_{q=1} = S_{int} - D_{q=1}$, that is, the 'net entropy' is equal to the difference between intrinsic entropy and disentropy. The intrinsic entropy in (16), $S_{int}$, is the term that contains the $q$-logarithm of the Lambert-Tsallis function. One should note that $D_q$ is the generalization of $D_{q=1}$ in the same way that Tsallis entropy is the generalization of Boltzmann-Gibbs (BG) entropy: The disentropy $D_{q=1}$, related to BG entropy, uses the Lambert function while $D_{q \neq 1}$, related to Tsallis entropy, uses the Lambert-Tsallis function. The disentropy related to Shannon entropy has $q = 1$ and it is obtained using eq. (4),

$$D = \sum_i p_i R_2(p_i). \quad (18)$$

One can also define the disentropy using the Lambert-Kaniadakis function [12]

$$D_\kappa = \sum_i p_i W_\kappa(p_i), \quad (19)$$

however, this disentropy does not come naturally from the relation between Kaniadakis logarithm and the Lambert-Kaniadakis function.

At last, one may also note that the minimal value of the disentropy is not zero. This happens because the Lambert function and its generalizations are equal to zero only when the argument is zero, hence, sometimes may be useful to use the normalized version of the disentropy: $D_{NORM} = [D - D_{min}]/[D_{max} - D_{min}]$, where $D_{max} = W_{q,K}(1)$ (or $R_\lambda(1)$) and $D_{min}$ is the disentropy of the uniform distribution. Thus, $0 \leq D_{NORM} \leq 1$.

In this direction, the present work shows theoretical developments and applications of the disentropy in different problems. Initially several concepts of classical and quantum information theory like Shannon's theorems and entanglement measures are rewritten using disentropy. Following, an equation for the black hole disentropy is provided. A protocol for image segmentation using disentropy is described and, at last, a disentropic distance measure between integer numbers is provided. This measure permits, for example, to measure the distance between a composed number and the set of prime powers.

## 2. Basic concepts of classical disentropy

Let the random variable $X = \{x_k \mid k = 1,2,...,K\}$ to represent the possible outcomes of an event (like a measurement). The result $x_k$ appears with probability $p_k$ where $p_k \geq 0$ and $\sum_{k=1}^{K} p_k = 1$. The amount of disinformation associated to $p_k$ is given by $d_q = W_q(p_k)$. One may note that $W_q(0) = 0$ and $W_q(x) > 0$ for $x > 0$ [10]. The average amount of disinformation is the disentropy

$$D_q(X) = \sum_{k=1}^{K} p_k^q W_q(p_k). \tag{20}$$

For a delta distribution one has $D_q = W_q(1)$, its maximal value, while for a uniform distribution $D_q = K^{(1-q)}W(K^{-1})$, its minimal value. Following eq. (20), the joint disentropy of the random variables $X$ and $Y$ is given by

$$D_q(X,Y) = \sum_{k=1, j=1}^{K,J} p^q(x_k, y_j) W_q(p(x_k, y_j)), \tag{21}$$

where $p(x_k, y_j)$ is the joint probability of $X = x_k$ and $Y = y_j$. The mutual disentropy, by its turn, is defined as

$$D_q(X:Y) = D_q(X) + D_q(Y) - D_q(X,Y). \tag{22}$$

Using the joint disentropy, the conditional disentropy can be defined in the traditional way as being

$$D_q(X|Y) = D_q(Y) - D_q(X,Y). \qquad (23)$$

The relative disentropy can be defined in four different ways

$$D_q(X\|Y) = \begin{cases} \sum_k p_k^q |W_q(p_k) - W_q(t_k)| & (24.a) \\ \sum_k p_k^q (W_q(p_k) - W_q(t_k)) & (24.b) \\ \sum_k p_k^q W_q(p_k - t_k) & (24.c) \\ \sum_k p_k^q W_q(|p_k - t_k|) & (24.d) \end{cases}.$$

In (24) the distribution $p_k$ ($t_k$) is associated to the possible values of the random variable $X$ ($Y$). Since $W_q(x)$ can return a real value for some negative input values, the relative disentropies (24.b) and (24.c) can also return a negative value.

At last, the disentropic uncertainty is written as

$$D_q(X) + D_q(Y) \leq 2W_q(1). \qquad (25)$$

Obviously, the value $2W_q(1)$ is not the tight bound value when $X$ and $Y$ are correlated.

## 3. Basic concepts of quantum disentropy

The quantum version of the disentropy is [10]

$$D_q(\rho) = \sum_n \lambda_n^q W_q(\lambda_n), \qquad (26)$$

where $\lambda_n$'s are the eigenvalues of the density matrix $\rho$. The quantum mutual disentropy is

$$D_q(\rho_A : \rho_B) = D_q(\rho_A) + D_q(\rho_B) - D_q(\Gamma_{AB}) \tag{27}$$

$$\rho_{A(B)} = Tr_{B(A)} \Gamma_{AB}. \tag{28}$$

In (28) $Tr$ is the partial trace operation. The quantum mutual disentropy is non-negative. The quantum conditional disentropy can be defined in the traditional way as being

$$D_q(A|B) = D_q(\Gamma_{AB}) - D_q(\rho_B). \tag{29}$$

The quantum conditional disentropy can be negative or positive for entangled states but it is negative for disentangled states. The quantum relative disentropy, by its turn, can be defined in four different ways

$$D_q(\rho\|\Gamma) = \begin{cases} \sum_n \lambda_n^q |W_q(\lambda_n) - W_q(\gamma_n)| & (30.a) \\ \sum_n \lambda_n^q (W_q(\lambda_n) - W_q(\gamma_n)) & (30.b) \\ \sum_n \lambda_n^q W_q(\lambda_n - \gamma_n) & (30.c) \\ \sum_n \lambda_n^q W_q(|\lambda_n - \gamma_n|) & (30.d) \end{cases}$$

where $\lambda_n$ and $\gamma_n$ are, respectively, the eigenvalues of the density matrices $\rho$ and $\Gamma$.

## 4. Disentropy in classical information theory

Initially, let us consider a memoryless source characterized by the alphabet $X = \{x_1, x_2, x_3, x_4,\ldots, x_{|X|}\}$ ($|X|$ is the cardinality of $X$) and the probabilities $\{p_1, p_2,\ldots,p_{|X|}\}$, where $p_i$ is the probability of the symbol $x_i$ to be emitted by the source. Consider now a sequence of $n$ symbols emitted by the source (every emission is an independent event): $X_n = x_{a1}, x_{a2}, x_{a3},\ldots,x_{an}$ ($a_1, a_2,\ldots a_n \in \{1, 2, 3\ldots |X|\}$). If the symbol $x_i$ appears $n_i$ times in $X_n$, then its relative frequency is $n_i/n$. The Shannon entropy and the corresponding disentropy of the sequence $X_n$ are given, respectively, by $\bar{H}(X_n) = -\sum_{i=1}^{|X|}(n_i/n)log_2(n_i/n)$ and $\bar{D} = \sum_{i=1}^{|X|}(n_i/n)R_2(n_i/n)$. The sequence $X_n$ is said to be $\delta$-typical if $|H(X) - \bar{H}(X_n)| \leq \delta$ or similarly $|D(X) - \bar{D}(X_n)| \leq \delta$, since

both conditions are satisfied simultaneously. That is, if the entropy of a sequence is close to the entropy of the source then the disentropy of that sequence is also close to the disentropy of the source (this happens because the relative frequency $n_i/n$ tends to the probability $p_i$ when $n$ grows). Hence, there exist a set of $\delta$-typical sequences, $T_\delta(X_n)$. It is well known that the probability of a sequence $X_n$ to belong to $T_\delta(X_n)$ is $1-\varepsilon$ where $\varepsilon$ goes to zero when $n$ goes to infinite. Furthermore, the cardinality of $T_\delta(X_n)$, for sufficiently large value of $n$, obeys the limits

$$(1-\varepsilon)2^{n[H(X)-\delta]} \leq |T_\delta(X_n)| \leq 2^{n[H(X)+\delta]}. \tag{31}$$

Thus, the entropy of the source $H(X)$ has a very clear operational meaning: for $n$ sufficiently large, the number of typical sequences is approximately $2^{nH(X)}$. On the other hand, one can check numerically that

$$H(X) + D(X)\log_2|X| \approx \log_2|X|, \tag{32}$$

where here $D(X)$ is the normalized version of the disentropy. For example, one can see in Fig. 3 $H(p) + D(p)$ for a binary source $\{x_0, x_1\}$ ($|X| = 2$) with distribution $\{p, 1-p\}$.

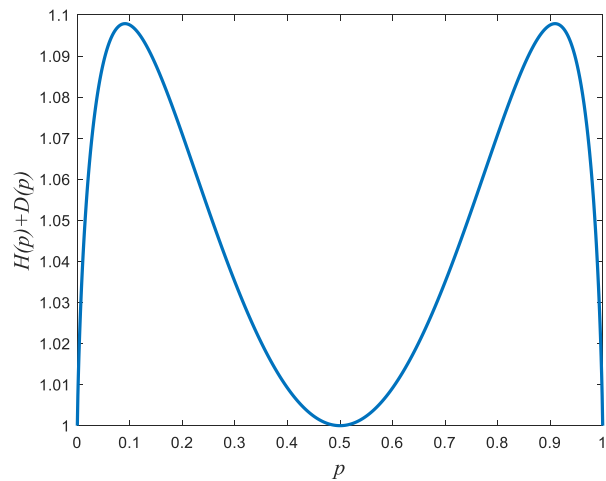

Fig. 3 – $H(p) + D(p)$ versus $p$.

Thus, one has for the total number of sequences with $n$ symbols:

$$|X_n|^n = 2^{n\log_2|X_n|} \approx 2^{n[H(X)+\log_2|X_n|D(X)]} = 2^{nH(X)}2^{n\log_2|X_n|D(X)} \approx |T_\delta(X_n)|2^{n\log_2|X_n|D(X)} \Rightarrow$$

$$2^{-n\log_2|X_n|D(X)} \approx \frac{|T_\delta(X_n)|}{|X_n|^n}. \quad (33)$$

Hence, the disentropy also has an operational meaning: $2^{-n\log_2|X|D(X)}$ $(1 - 2^{-n\log_2|X|D(X)})$ is the fraction of typical (non-typical) sequences. For a source with uniform distribution one has $D(X) = 0$ and $|X_n|^n = |T_\delta(X_n)|$, as expected. On the other hand, for an almost deterministic source $D(X) \approx 1$ and, hence, the fraction of typical sequences is almost zero.

The Shannon's source coding theorem states that $H(X) \leq \langle L \rangle \leq H(X)+1$ bits, where $H(X)$ is the Shannon entropy of the used alphabet $X$ with $r$ symbols. Each symbol is selected by the source with probability $p_i$, $i = 1,\ldots,r$. Furthermore, $\langle L \rangle = \sum_{i=1}^{r} p_i l_i$, where $l_i$ is the number of bits of the $i$-th codeword. Similarly, using eq. (32) the Shannon's source coding theorem can be rewritten using the disentropy:

$$H(X) \leq \langle L \rangle \leq H(X)+1 \Rightarrow D(X) - \frac{1}{\log_2|X|} \leq 1 - \frac{\langle L \rangle}{\log_2|X|} \leq D(X) \Rightarrow D(X) - \frac{1}{\log_2|X|} \leq \langle \Lambda \rangle \leq D(X), \quad (34)$$

Using (32) one can also get the following inequality

$$\lceil \log_2|X| \rceil [1 - D(X)] \leq \langle L \rangle \leq H(X)+1, \quad (35)$$

where $\lceil x \rceil$ means the smallest integer number larger than $x$, with the following interpretation: the best code has $\langle L \rangle \sim H(X)$ while the worst code has $\langle L \rangle \sim \lceil \log_2|X| \rceil [1 - D(X)]$. For example, let us consider the alphabet $X = \{x_1, x_2, x_3, x_4, x_5\}$ having the following probabilities $\{0.4, 0.2, 0.15, 0.15, 0.1\}$. A good code ($C_1$) and a bad code ($C_2$) are shown in Table I.

Table I – Good code ($C_1$) and bad code ($C_2$) for the alphabet $X = \{x_1, x_2, x_3, x_4, x_5\}$.

| Symbol | Probability | Codeword 1 | Codeword 2 |
|---|---|---|---|
| $x_1$ | 0.4 | 1 | 011 |
| $x_2$ | 0.2 | 000 | 000 |
| $x_3$ | 0.15 | 001 | 001 |
| $x_4$ | 0.15 | 010 | 010 |
| $x_5$ | 0.1 | 011 | 10 |

It can be easily shown that $\langle L_1 \rangle = 2.2$, $H(X) \sim 2.1464$; $\langle \Lambda \rangle \sim 0.0525$, $D(X) \sim 0.07951$; $\langle L_2 \rangle = 2.9$, $[log_2(5)][1 - D(X)] \sim 2.7614$.

Now, consider a noisy channel modelled by the conditional probability $p(y|x)$, as shown in Fig. 4.

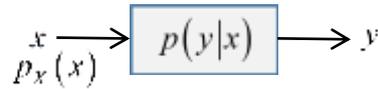

Fig. 4 – Classical noisy channel modelled by the conditional probability $p(y|x)$.

The input of the noisy channel is the random variable $X$ with probability distribution $p_X(x)$. The output random variable is $Y$. The Shannon-Hartley's theorem states that, in order to avoid errors, the transmission rate $R$ must be smaller than the channel's capacity given by $C = sup_{p_{X(x)}} I(X:Y)$, where $I(X:Y)$ is the mutual information. Similarly, one can define the channel capacity based on disentropy as being

$$C_q = \underset{p_X(x)}{Inf} \, D_q\left(X:Y\right). \tag{36}$$

For a given channel, the probability distribution $p_X(x)$ that maximizes $C$ is the same that minimizes $C_q$. As an example, let us consider a binary discrete memoryless channel. The source symbols and probabilities are $\{('0', p_0), ('1', 1-p_0)\}$ while the channel transition probability is $p_c$. For this channel one has the following channel capacity, reached when $p_0 = 1/2$,

$$C = 1 + \left(1 - p_c\right)\log_2\left(1 - p_c\right) + p_c \log_2\left(p_c\right). \tag{37}$$

In Fig. 5 one can see the plot of $D_q$ versus $p_0$ for different values of $p_c$ and $q$.

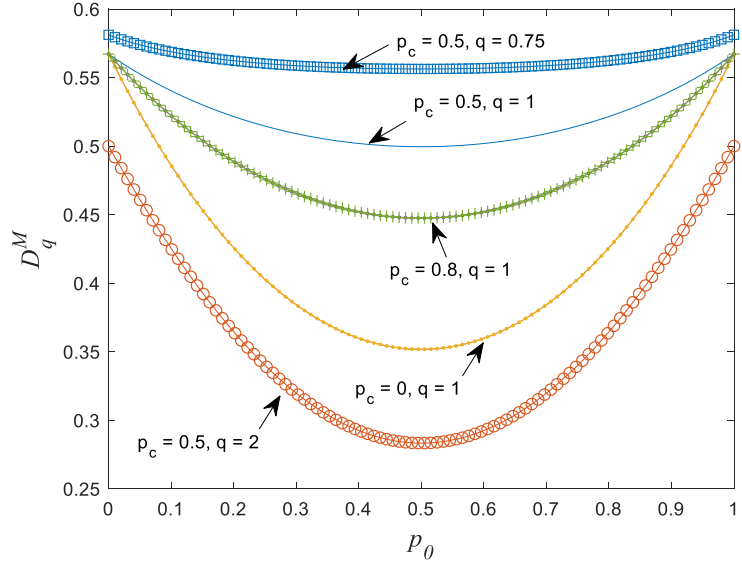

Fig. 5 - Mutual disentropy versus $p_0$ for different values of $p_c$ and $q$.

As can be seen in Fig. 5, the minimal value of the mutual disentropy occurs always for $p_0 = \frac{1}{2}$, as expected. The value of $C_q$ is

$$C_q = \frac{4}{2^q}W_q\left(\frac{1}{2}\right) - \frac{(1-p_c)^q}{2^{q-1}}W_q\left(\frac{1-p_c}{2}\right) - \frac{p_c^q}{2^{q-1}}W_q\left(\frac{p_c}{2}\right). \tag{38}$$

In particular, for $q = 1$ one has

$$C_1 = 2W\left(\frac{1}{2}\right) - (1-p_c)W\left(\frac{1-p_c}{2}\right) - p_c W\left(\frac{p_c}{2}\right). \tag{39}$$

Now, let us discuss the classical Fano inequality. Initially, $X$ is a random variable whose values are chosen randomly from the alphabet $\chi = \{x_1, x_2, x_3...x_n\}$. It is the input of a classical channel while the random variable $Y$, an estimation of $X$, is the output. The classical Fano inequality states that

$$H(X|Y) \leq H(e) + p_e \log(|\chi| - 1) \tag{40}$$

$$H(e) = -p_e \log(p_e) - (1-p_e)\log(1-p_e). \tag{41}$$

In (29) $p_e$ is the probability of $Y \neq X$ and $|\chi|$ is the cardinality of $\chi$. The classical Fano inequality using disentropy is as follow

$$D_q(X|Y) \leq D_q(e) + p_e^q W_q(|\chi|-1) \tag{42}$$

$$D_q(e) = p_e^q W_q(p_e) + (1-p_e)^q W_q(1-p_e). \tag{43}$$

As an example, let us consider again the binary discrete memoryless channel used before. In this case, $p_e = p_c$. Thus, for $p_0 = \frac{1}{2}$, one gets

$$D_q(X|Y) = 2\left(\frac{1}{2}\right)^q W_q\left(\frac{1}{2}\right) - 2\left[\frac{(1-p_c)}{2}\right]^q W_q\left(\frac{(1-p_c)}{2}\right) - 2\left(\frac{p_c}{2}\right)^q W_q\left(\frac{p_c}{2}\right) \tag{44}$$

$$D_q(e) + p_e^q W_q(|\chi|-1) = p_c^q W_q(p_c) + (1-p_c)^q W_q(1-p_c) + p_c^q W_q(1). \tag{45}$$

Figure 6 shows eqs. (44) and (45) versus $p_c$ for three different values of $q$ (0.75, 1, 1.25). As it can be noted in Fig. 3, the Fano inequality is always obeyed.

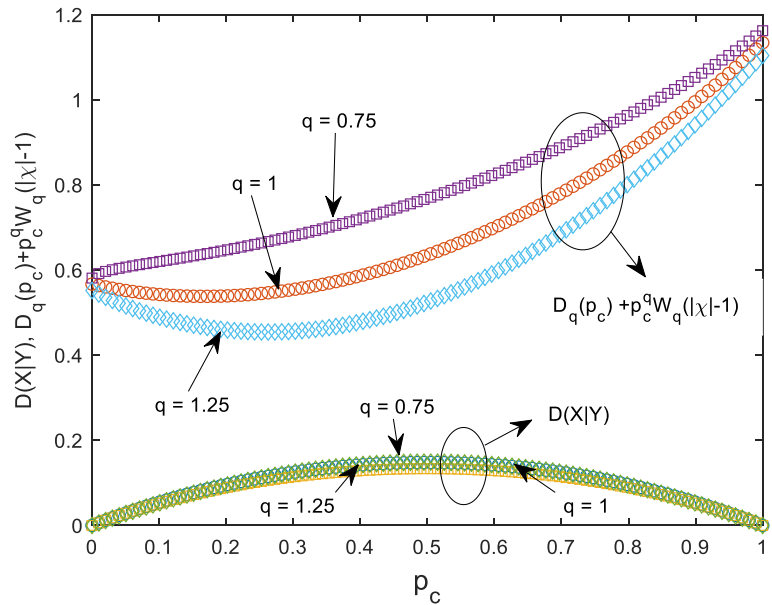

Fig. 6 – $D_q(X|Y)$ and $D_q(p_c) + p_c^q W_q(|\chi|-1)$ versus $p_c$ ($p_0 = 0.5$) for three different values of $q$ (0.75, 1, 1.25).

## 5. Disentropy in quantum information theory

One of the most important theorems in quantum information is the Holevo's theorem:

$$I(A:B) \leq S\left(\sum_i p_i \rho_i\right) - \sum_i p_i S(\rho_i). \tag{46}$$

In (46) $S(\rho)$ is the von Neumann quantum entropy. The single-use Holevo-Schumacher-Westmoreland capacity of a given quantum channel is [14]

$$C = \max_{p_i, \rho_i} \left\{ S\left(\mathcal{N}\left(\sum_i p_i \rho_i\right)\right) - \sum_i p_i S(\mathcal{N}(\rho_i)) \right\}. \tag{47}$$

where $\mathcal{N}$ is the trace-preserving map that models the channel, $Tr(\mathcal{N}(\rho)) = Tr(\rho)$. The theorem similar to Holevo's one using the disentropy is

$$D_q(X:Y) \leq \sum_i p_i D_q(\rho_i) - D_q\left(\sum_i p_i \rho_i\right). \tag{48}$$

Furthermore, the set $\{p_i, \rho_i\}$ that maximizes (47) is the same one the minimizes

$$C_q = \min_{p_i, \rho_i} \sum_i p_i D_q(\mathcal{N}(\rho_i)) - D_q\left(\mathcal{N}\left(\sum_i p_i \rho_i\right)\right). \tag{49}$$

Consider, now, $|\psi_{AB}\rangle \in H_{AB} = H_A \otimes H_B$, with $\dim(H_A) = \dim(H_B) = d \geq 2$. The system $A$ experiences a completely positive and trace-preserving quantum operation $\varepsilon$ while system $B$ is let isolated. The final state after such interaction is $\rho_o = (\varepsilon \otimes I)|\psi_{AB}\rangle\langle\psi_{AB}|(\varepsilon \otimes I)^\dagger$, where $I$ is the identity operator. The entanglement fidelity, as defined by Schumacher, is $F(\varepsilon) = \langle\psi_{AB}|\rho_o|\psi_{AB}\rangle$ [15], while the von Neumann entropy exchange is $S(\rho_o)$. The quantum Fano inequality (QFI) is, hence, given by

$$S(\rho_o) \leq H(F(\varepsilon)) + [1 - F(\varepsilon)] \log(d^2 - 1). \tag{50}$$

Without proof I conjecture that the QFI using disentropy is given by

$$D_q(\rho_o) \leq D_q(F(\varepsilon)) + [1 - F(\varepsilon)]^q W_q(d^2 - 1). \tag{51}$$

Now, let us discuss some quantum correlation measures based on disentropy. It was shown that the quantum disentropy can be used to measure the disentanglement of bipartite of qubit states: The disentanglement measure of the two-qubit state $|\varphi\rangle_{AB} = \sqrt{\lambda_0}|00\rangle_{AB} + \sqrt{\lambda_1}|11\rangle_{AB}$ (here $\lambda_0$ and $\lambda_1$ are the Schmidt coefficients and $\lambda_0 + \lambda_1 = 1$) is given by the quantum disentropy of the partial state $\rho_A = \text{Tr}_B(|\varphi\rangle_{AB})$ ($\rho_B$ could also be used) [10]

$$D_E \equiv D_q(\rho_A) = \lambda_0^q W_q(\lambda_0) + \lambda_1^q W_q(\lambda_1). \tag{52}$$

From (52) it is easy to note the disentanglement is maximal when $\rho_A$ is a pure state ($\lambda_0 \lambda_1 = 0$) and minimal when $\rho_A$ is the maximally mixed state ($\lambda_0 = \lambda_1 = 1/2$). The disentanglement of a mixed two-qubit state can be calculated using

$$D_E(\rho) = \min \sum_i p_i \, _Q D_q(\rho_A^i) \bigg| \rho = \sum_i p_i \sigma_i \tag{53}$$

$$\rho_A^i = Tr_B(\sigma_i), \tag{54}$$

or using the distance between quantum states, where the relative quantum disentropy is the distance measure:

$$D_E(\rho) = \min_{\Gamma \in E} D_q(\rho \| \Gamma). \tag{55}$$

In eq. (55) $E$ is the set containing all entangled states. However, using the concurrence $C(\rho)$ [16], the disentanglement of a two-qubit state is given by

$$D_E(\rho) = \left(\frac{1}{2} + \frac{\sqrt{1-C(\rho)}}{2}\right)^q W_q\left(\frac{1}{2} + \frac{\sqrt{1-C(\rho)}}{2}\right) + \left(\frac{1}{2} - \frac{\sqrt{1-C(\rho)}}{2}\right)^q W_q\left(\frac{1}{2} - \frac{\sqrt{1-C(\rho)}}{2}\right). \quad (56)$$

At last, one should note that, like entanglement, the disentanglement also obeys some monogamy inequalities For example, for a three qubit state one has [10]

$$D_E(\rho_{A\_BC}) \leq D_E(\rho_{AB}) + D_E(\rho_{AC}) \quad (57)$$

$$D_E(\rho_{B\_AC}) \leq D_E(\rho_{AB}) + D_E(\rho_{BC}) \quad (58)$$

$$D_E(\rho_{C\_AB}) \leq D_E(\rho_{AC}) + D_E(\rho_{BC}). \quad (59)$$

Furthermore, the disentanglement measure for pure three qubit GHZ-class states is [10]

$$D_E(\rho) = \left(\frac{1}{2} + \frac{\sqrt{1-\tau_3(\rho)}}{2}\right)^q W_q\left(\frac{1}{2} + \frac{\sqrt{1-\tau_3(\rho)}}{2}\right) + \left(\frac{1}{2} - \frac{\sqrt{1-\tau_3(\rho)}}{2}\right)^q W_q\left(\frac{1}{2} - \frac{\sqrt{1-\tau_3(\rho)}}{2}\right), \quad (60)$$

where $\tau_3$ is the three-tangle introduced in [17].

The quantum disentropy can also be used to analyse the security of quantum protocols. For example, a quantum key distribution protocol (QKD) is considered secure if the mutual information between Alice and Bob is larger than the mutual information between Alice and Eve: $I_{AB} > I_{AE}$. Similarly, under the same conditions, a QKD protocol can be considered secure if $D_q(A:B) < D_q(A:E)$. For example, the key rate of GLLP [18,19] is given by

$$R \geq \sigma\left\{-Q_\mu H_2(E_\mu) + Q_1[1 - H_2(e_1)]\right\}. \quad (61)$$

In eq. (61) $\sigma = 1/2$ for BB84 protocol. When Alice uses a coherent state with mean photon number $\mu$, $Q_\mu$ is the probability for Bob to get a detection in a pulse that Alice and Bob use the same basis, $E_\mu$ is the probability for Bob to get a wrong detection in a pulse that Alice and Bob use the same basis. The variables $Q_1$ ($e_1$) has the same meaning as $Q_\mu$ ($E_\mu$) when Alice uses a single-photon pulse. At last, $H_2(x) = -x\log_2(x)-(1-x)\log_2(1-x)$. Using (32) to rewrite (61) using the disentropy one immediately gets

$$R \geq q\left\{-Q_\mu\left[1-D(E_\mu)\right]+Q_1 D(e_1)\right\}. \tag{62}$$

where $D(x) = [xR_2(x)+(1-x)R_2(1-x)- R_2(1/2)]/ [R_2(1)- R_2(1/2)]$.

At last, we discuss the quantum discord using the disentropy. Let $\rho_{AB}$ be a bipartite state and $\Pi_j^B$ a set of measurements on subsystem $B$. The reduced state of subsystem $A$ after the measurement is given by [20]

$$\rho_A^j = \frac{1}{p_j} Tr_B\left[\left(1_A \otimes \Pi_j^B\right)\rho_{AB}\left(1_A \otimes \Pi_j^B\right)\right] \tag{63}$$

$$p_j = Tr_B\left[\left(1_A \otimes \Pi_j^B\right)\rho_{AB}\right]. \tag{64}$$

The conditional disentropy of subsystem $A$ for a known subsystem $B$ is

$$D_q(A|B) \equiv \sum_j p_j D_q\left(\rho_A^j\right) \tag{65}$$

$$D_q^J(A:B) = D_q(A) - D_q(A|B). \tag{66}$$

At last, the quantum discord using disentropy is written as

$$D_q^{dsc} = D_q(A:B) - \max_{\{\Pi_j^B\}} D_q^J(A:B). \tag{67}$$

## 6. Disentropy of the Wigner function

At a first glance, it seems that disentropy does not bring something really new. It seems to be just an opposite point of view: order instead of disorder or disentanglement instead of entanglement. However, this is not true. The Lambert function and its generalizations can assume real values for a range of negative values of the argument.

For example, $W_1(z)$ is real for $z > -1/\exp(1)$ while $W_2(z)$ is real for $z > -1$. This means the disentropy of some functions with negative values can be calculated. Thus, the disentropy of some quasi-probability distributions can be calculated. If $w(x,y)$ is the Wigner function of a given quantum state, the disentropy and relative disentropy of the Wigner function are given by [21]

$$D_q = \iint_{y\,x} w^q(x,y) W_q(w(x,y)) dx dy \tag{68}$$

$$D_q(w(x,y) \| w_{ref}(x,y)) = \iint_{y\,x} w^q(x,y) W_q(w(x,y) \| w_{ref}(x,y)) dx dy, \tag{69}$$

where

$$W_q(w(x,y) \| w_{ref}(x,y)) = \begin{cases} |W_q(w(x,y)) - W_q(w_{ref}(x,y))| & (70.a) \\ W_q(w(x,y)) - W_q(w_{ref}(x,y)) & (70.b) \\ W_q(w(x,y) - w_{ref}(x,y)) & (70.c) \\ W_q(|w(x,y) - w_{ref}(x,y)|). & (70.d) \end{cases}$$

The disentropy of the Wigner function has some interesting properties. If the Wigner function is totally positive its disentropy is also positive and it measures the order or the purity of the quantum state. On the other hand, if the Wigner function has negative parts, its disentropy can positive or negative. When it is negative, the disentropy can work as quantumness measure [21]. When the Wigner function has negative parts but its disentropy is positive, there is a contest between quantumness and order measures. As an example, let us consider the following mixed state

$$\rho = e^{-t} |\xi\rangle^{(s)(s)} \langle\xi| + (1 - e^{-t}) |00\rangle\langle00|, \tag{71}$$

Where $|\xi\rangle^{(s)}$ is the vortex state described in [22]

$$|\xi\rangle^{(s)} = \frac{e^{i\varphi}}{\cosh(r)} a^\dagger |\xi\rangle = \frac{e^{i\varphi}}{\cosh^2(r)} \sum_{n=0}^{\infty} e^{in\varphi} \tanh^n(r) \sqrt{n+1} |n+1, n\rangle \tag{72}$$

and $|\xi\rangle$ is the two-mode squeezed vacuum state

$$|\xi\rangle = \exp\left(\xi a^{\dagger}b^{\dagger} - \xi^* ab\right)|00\rangle. \tag{73}$$

In (73) $\xi = re^{i\varphi}$. The Wigner function of the state (71) is

$$w(\lambda, \beta) = \frac{4}{\pi^2}\left(1 - e^{-t}\right)\exp\left[-2\left(|\beta|^2 + |\lambda|^2\right)\right] + e^{-t}\frac{4}{\pi^2}\left[4\left|\cosh(r)\lambda - \sinh(r)e^{i\varphi}\beta^*\right|^2 - 1\right] \times \\ \exp\left[-2\left(\left|\cosh(r)\lambda - \sinh(r)e^{i\varphi}\beta^*\right|^2 + \left|\cosh(r)\beta^* - \sinh(r)\lambda e^{-i\varphi}\right|^2\right)\right]. \tag{74}$$

In (74) $\lambda = x_1 + iy_1$ and $\beta = x_2 + iy_2$ where $x_1$ and $y_1$ ($x_2$ and $y_2$) are the quadratures of the first (second) mode. The curve of the disentropy ($q = 2$) of (74) versus $t$ can be seen in Fig. 7.

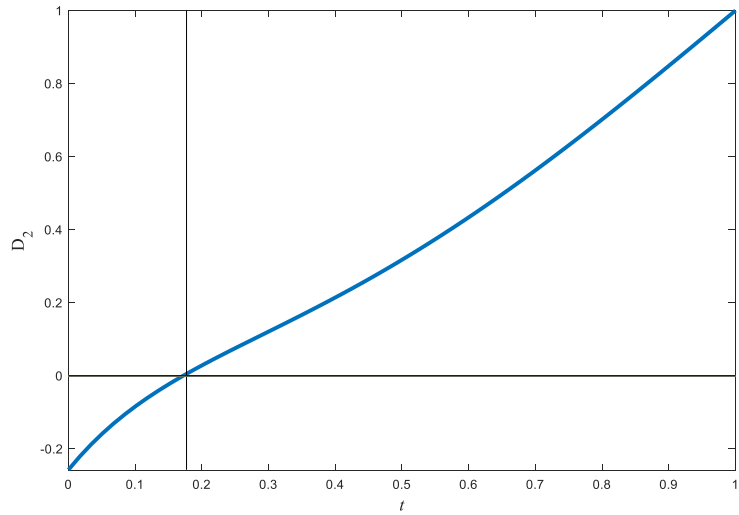

Fig. 7 – Normalized disentropy of eq. (74) versus $t$.

In Fig. 7, $D_2 = 0$ for $t \sim 0.178$. The effect of the added photon can be seen in the negative part of the disentropy and it decreases when $t$ increases (the vacuum component predominates).

Now, let us consider an initially coherent state propagating in a Kerr medium. It evolves according to [23,24]

$$|\psi(\tau;\beta)\rangle = U(\tau)\left|\beta e^{-\tau/\sigma}\right\rangle = e^{-i\frac{H}{\hbar}t}\left|\beta e^{-\tau/\sigma}\right\rangle = e^{-i\frac{\kappa t}{2}a^\dagger a^\dagger aa}\left|\beta e^{-\tau/\sigma}\right\rangle = e^{-\frac{\left|\beta e^{-\tau/\sigma}\right|^2}{2}}\sum_{n=0}^{\infty}\frac{\left(\beta e^{-\tau/\sigma}\right)^n}{\sqrt{n!}}e^{-i\frac{\tau}{2}n(n-1)}|n\rangle. \quad (75)$$

The unitless time is $\tau = \kappa t$ where $\kappa$ is a nonlinear constant proportional to the third order susceptibility $\chi(3)$. The real exponential $\exp(-\tau/\sigma)$ was included to break the periodicity of the quantum state (75). Without the real exponential it would be periodic with period $2\pi$: $|\psi(\tau;\beta)\rangle = |\psi(\tau + 2\pi;\beta)\rangle$. The corresponding time varying Wigner function of the state (75) is [24]

$$w(\lambda) = \frac{2}{\pi}e^{-2|\lambda|^2}e^{-|\beta\exp(-\tau/\sigma)|^2} \times \\ \sum_{n=0}^{\infty}\frac{\left(2\beta^*\exp(-\tau/\sigma)\lambda e^{i\frac{\tau}{2}}\right)^n}{n!}e^{-i\frac{\tau}{2}n^2}\sum_{m=0}^{\infty}\frac{\left(2\beta\exp(-\tau/\sigma)\lambda^* e^{-i\frac{\tau}{2}}\right)^m}{m!}e^{i\frac{\tau}{2}m^2}e^{-|\beta\exp(-\tau/\sigma)|^2 e^{i\tau(m-n)}}. \quad (76)$$

Using $\beta = 2$ and $\sigma = 8\pi$, one can see in Fig. 8 the disentropy $D_2(w(\lambda))$ versus $\tau$ in the interval $[0,2\pi]$. The break of the periodicity of the quantum state in eq. (75) by the real exponential is clearly seen in Fig. 8.

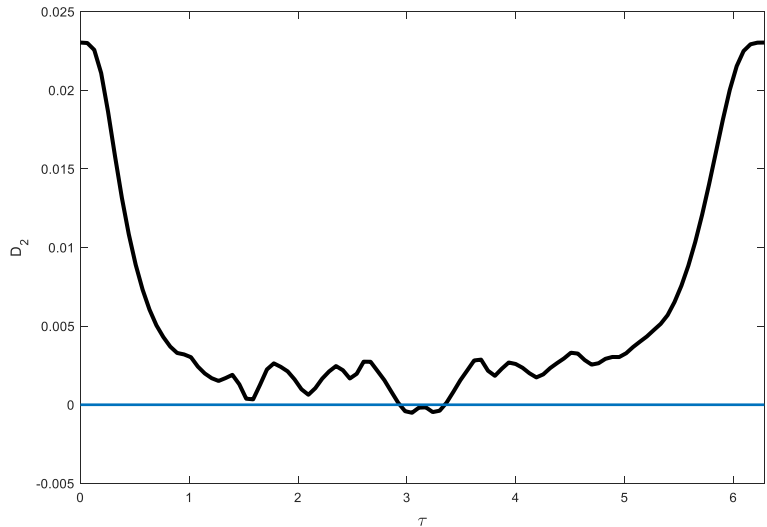

Fig. 8 – Disentropy ($q = 2$) of the Wigner function in eq. (76) versus $\tau$.

Since the state in eq. (75) is always pure, the disentropy measures its quantumness that is maximal when a superposition of coherent states (cat states) is produced. This is seen in the negative part of the disentropy in Fig. 8 [21].

In the calculation of the disentropy of the Wigner function, one has to be careful with the value of $q$ used. As one can see in eqs. (8)-(10), for example, the Lambert-Tsallis function with a fractional $q$ value has fractional powers, what can result in a complex value for the disentropy when the Wigner function is negative. In order to avoid this problem, one can use another expression for the disentropy and relative disentropy of the Wigner function [11]

$$D_{q,\alpha} = W_q\left(\int w^\alpha(x)\,dx\right) \tag{77}$$

$$D_{q,\alpha}(w_1\|w_2) = W_q(1) - W_q\left(\iint_{x,y} \left[w_2^{\frac{1-\alpha}{2\alpha}}(x,y)\,w_1(x,y)\,w_2^{\frac{1-\alpha}{2\alpha}}(x,y)\right]^\alpha dxdy\right) \quad \alpha \in [1/2,1)\cup(1,\infty) \tag{78}$$

$$D_{q,\alpha}(w_1\|w_2) = W_q(1) - W_q\left(\iint_{x,y} w_1^\alpha(x,y)\,w_2^{1-\alpha}(x,y)\,dxdy\right) \quad \alpha \in [0,1)\cup(1,2). \tag{79}$$

In this case, the disentropy of the Wigner function cannot be used to measure the quantumness, but it can be used to measure the disentanglement. Consider, for example, the two-mode squeezed vacuum state (73) whose Wigner function is

$$w(\lambda,\beta) = \frac{4}{\pi^2}\exp\left[-2\left(\left|\cosh(r)\lambda - \sinh(r)e^{i\varphi}\beta^*\right|^2 + \left|\cosh(r)\beta^* - \sinh(r)\lambda e^{-i\varphi}\right|^2\right)\right]. \tag{80}$$

In (80) $\lambda = x_1 + iy_1$ and $\beta = x_2 + iy_2$ where once more $x_1$ and $y_1$ ($x_2$ and $y_2$) are the quadratures of the first (second) mode. In Fig. 9 one can see the plot of $D_{q=3/2,\alpha=2}$ versus $r$.

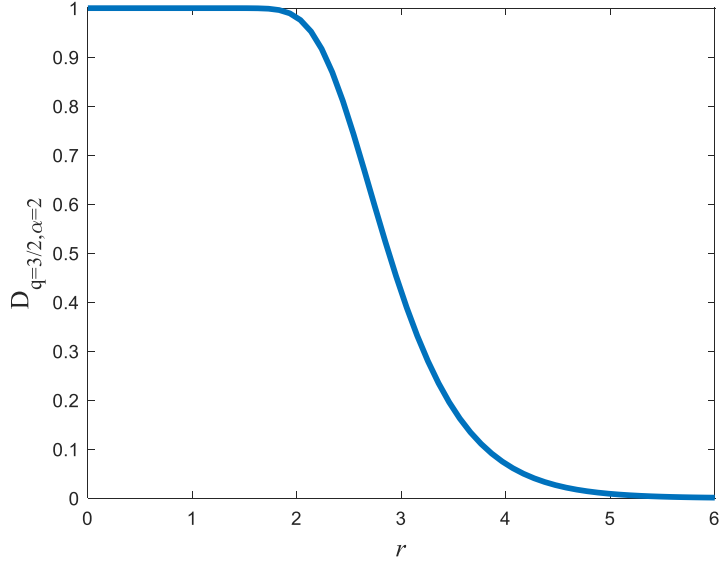

Fig. 9 – Normalized disentropy ($D_{q,\alpha}$) of the Wigner function (80) versus $r$ using $q = 3/2$ and $\alpha = 2$.

The initial state ($r = 0$) is the disentangled state $|0\rangle|0\rangle$, hence the disentanglement mesured by the disentropy $D_{q=3/2,\alpha=2}$ is maximal at this point. The entanglement increases when $r$ increases, therefore the disentanglement measured by $D_{q=3/2,\alpha=2}$ decreases when $r$ increases.

## 7. Non-Extensive Statistical Mechanics and Black Hole Disentropy

The loop quantum gravity provides an estimation of the microstates of a black hole with classical horizon area $A$, permitting the calculation of its entropy. The result is $S_{bh} = (\lambda_0/\gamma)A$ where $\lambda_0$ is a numerical constant resulting from the underlying statistical mechanics while $\gamma$ is the Barbero–Immirzi parameter [25]. However, according to Bekenstein–Hawking area law, the (maximal) entropy of a black hole is $S_{bh} = A/4l_p^2$ where $l_p$ is the Planck's length [26,27]. Here, the logarithmic correction due to quantum fluctuations that decrease the entropy is not considered [28]. This forces a unique relation between $\lambda_0/\gamma$ what imposes a value to $\gamma$. One way to overcome this problem is to consider a non-extensive statistical model, for example, using Tsallis' entropy instead of Boltzmann-Gibbs' entropy. In this case one has $S_q = \ln_q(\Omega) = A/4l_p^2$, where $\Omega$ is the total number of microstates of the system under consideration. Thus, it is possible to get

a relation between $q$ and $\gamma$ in such way that Bekenstein–Hawking area law is always satisfied [25]:

$$\gamma \ln\left[1+\frac{A}{4l_p^2}(1-q)\right] = \frac{\ln(2)}{\pi\sqrt{3}}\frac{A}{4l_p^2}(1-q). \tag{81}$$

Hence, if one can experimentally measure the value of $\gamma$, $\gamma_{exp}$, the correct value of $q$ can be obtained from (81),

$$q = 1 + \frac{4l_p^2}{A}\left[1+\frac{\pi\sqrt{3}\gamma_{exp}}{\ln(2)}W\left(-\frac{\ln(2)}{\gamma_{exp}\pi\sqrt{3}}e^{-\frac{\ln(2)}{\gamma_{exp}\pi\sqrt{3}}}\right)\right]. \tag{82}$$

Note that one must have $\gamma_{exp} < \ln(2)/(\pi\sqrt{3})$ otherwise eq. (82) will provide $q = 1$, the obvious result that can be obtained directly from (81): Any value of $\gamma$ is possible if $q = 1$. On the other hand, since $\Omega = \exp_q(A/4l_p^2)$, the disentropy of the black hole is given by

$$D_q(\Omega) = \sum_{i=1}^{\Omega}\frac{1}{\Omega^q}W_q\left(\frac{1}{\Omega}\right) = \Omega^{1-q}W_q\left(\frac{1}{\Omega}\right) = \left[1+(1-q)\frac{A}{4l_p^2}\right]W_q\left(e_q^{(-A/4l_p^2)/\left(1+(1-q)\frac{A}{4l_p^2}\right)}\right). \tag{83}$$

In (83) the value of $q$ used is that one obtained in (82). As one can see in eqs. (82)-(83), when $A$ grows, $q$ tends to 1 (Boltzmann-Gibbs' statistics) and $D_q$ tends to zero ($W_q(0) = 0$ for any $q$). In fact, for $q = 1$ one has $D = W(\exp(-A/4l_p^2))$. As expected, the larger the entropy the lower is the disentropy. The disentropy found in [12] using Kaniadakis statistics is

$$D_\kappa = W_\kappa\left[\left(\frac{\kappa A}{4l_p^2}+\sqrt{\frac{\kappa^2 A^2}{16l_p^4}+1}\right)^{-\frac{1}{\kappa}}\right], \tag{84}$$

whose limit value when the Boltmann-Gibbs statistics is recovered is also $D_{\kappa=0} = W(\exp(-A/4l_p^2))$.

## 8. The Lambert-Tsallis Operator: $A\exp_q(A) = B$

In the present section, we are interested in solving the following equation

$$\hat{A}e_q^{\hat{A}} = \hat{B}, \tag{85}$$

where $\hat{A}e_q^{\hat{A}}$ is the Lambert-Tsallis operator. In (85) $\hat{A}$ and $\hat{B}$ are quantum operators or square matrices. The goal is to find out the conditions required by $\hat{A}$ and $\hat{B}$ in order to eq. (85) to have solution.

Initially, we assume that $\hat{A}$ and $\hat{B}$ are Hermitean operators ($\hat{A} = \hat{A}^\dagger$ and $\hat{B} = \hat{B}^\dagger$). Using the spectral decomposition of $\hat{A}$, one has

$$\begin{aligned} e_q^{\hat{A}} &= \left[I + (1-q)\hat{A}\right]^{1/(1-q)} = \left[\sum_n |a_n\rangle\langle a_n| + (1-q)\sum_n a_n |a_n\rangle\langle a_n|\right]^{1/(1-q)} = \\ &\left[\sum_n |a_n\rangle\langle a_n| + (1-q)a_n|a_n\rangle\langle a_n|\right]^{1/(1-q)} = \left[\sum_n |a_n\rangle\langle a_n|[1+(1-q)a_n]|a_n\rangle\langle a_n|\right]^{1/(1-q)} \end{aligned} \tag{86}$$

In (86) $I$ is the identity operator and $a_n$ and $|a_n\rangle$ are, respectively, the eigenvalues and eigenstates of $\hat{A}$. For $1/(1-q) \in \mathbb{Z}$, the last term in the right hand side of (86) is

$$e_q^{\hat{A}} = \left[\sum_n |a_n\rangle\langle a_n|[1+(1-q)a_n]|a_n\rangle\langle a_n|\right]^{\frac{1}{1-q}} = \sum_n [1+(1-q)a_n]^{\frac{1}{1-q}}|a_n\rangle\langle a_n| = \sum_n e_q^{a_n}|a_n\rangle\langle a_n|. \tag{87}$$

Now, using (87) in (85) and the spectral decompositions of $\hat{B}$, one gets

$$\hat{A}e_q^{\hat{A}} = \hat{B} \Rightarrow \sum_m a_m|a_m\rangle\langle a_m|\sum_n e_q^{a_n}|a_n\rangle\langle a_n| = \sum_n b_n|b_n\rangle\langle b_n| \Rightarrow \\ \sum_n a_n e_q^{a_n}|a_n\rangle\langle a_n| = \sum_n b_n|b_n\rangle\langle b_n| \tag{88}$$

If $[\hat{A}, \hat{B}] = \hat{A}\hat{B} - \hat{B}\hat{A} = 0$, then $|a_n\rangle = |b_n\rangle$ for all $n$ and eq. (88) reduces to

$$\hat{A} e_q^{\hat{A}} = \hat{B} \Rightarrow \sum_n a_n e_q^{a_n} |a_n\rangle\langle a_n| = \sum_n b_n |a_n\rangle\langle a_n| \Rightarrow a_n e_q^{a_n} = b_n \Rightarrow a_n = W_q(b_n). \quad (89)$$

In other words, the eigenvalues of $\hat{A}$ and $\hat{B}$ are related by the Lambert-Tsallis function.

Equation (89) can be used in two cases. In the first one, the operator $\hat{B}$ is known and one wants to know which operator $\hat{A}$ satisfies eq. (85). For example, let us assume $\hat{B}$ to be the photon number operator $\hat{N}$ and $q = 2$. In these case, one has $b_n = n$ ($n \in \mathbb{N}$) and $a_n = W_2(n)$. Hence

$$\hat{A} e_2^{\hat{A}} = \hat{N} = \sum_n n |n\rangle\langle n| \Rightarrow \hat{A} = \sum_n W_2(n) |n\rangle\langle n| = \sum_n \frac{n}{n+1} |n\rangle\langle n|. \quad (90)$$

The second case is, knowing the operator $\hat{A}$, one wants to know $\hat{B}$. For example, let us assume that $\hat{A}$ is the photon number operator $\hat{N}$ and $q = 1/2$. In these cases, one has

$$\hat{N} e_{1/2}^{\hat{N}} = \sum_n n e_{1/2}^n |n\rangle\langle n| = \sum_n n(1+n/2)^2 |n\rangle\langle n|. \quad (91)$$

Now let us consider the finite dimensional case, thus, we use matrices instead of quantum operators. The equation to be solved is

$$A e_q^A = B, \quad (92)$$

where $A$ and $B$ are Hermitean matrices. Using the spectral decomposition of matrix $A$, one has

$$e_q^A = \left[ I + (1-q) A \right]^{1/(1-q)} = \left[ I + (1-q) U_A D_A U_A^\dagger \right]^{1/(1-q)} = \left\{ U_A \left[ I + (1-q) D_A \right] U_A^\dagger \right\}^{1/(1-q)}. \quad (93)$$

In (93) $U_A$ is the unitary matrix whose columns are the eigenvectors of $A$ and $D_A$ is the diagonal matrix whose entries are the eigenvalues of $A$. For $1/(1-q) \in \mathbb{Z}$, one gets

$$e_q^A = U_A \left[ I_n + (1-q) D_A \right]^{1/(1-q)} U_A^\dagger = U_A e_q^{D_A} U_A^\dagger. \tag{94}$$

Therefore

$$A e_q^A = B \Rightarrow \left( U_A D_A U_A^\dagger \right) \left( U_A e_q^{D_A} U_A^\dagger \right) = U_A D_A e_q^{D_A} U_A^\dagger = B \Rightarrow D_A e_q^{D_A} = U_A^\dagger B U_A. \tag{95}$$

Once again, if $[A,B] = 0$, then $B = U_A D_B U_A^\dagger$. Substituting in (95) one gets

$$D_A e_q^{D_A} = D_B \Rightarrow \begin{bmatrix} a_1 & 0 & 0 & 0 \\ 0 & a_2 & 0 & 0 \\ 0 & 0 & \ddots & 0 \\ 0 & 0 & 0 & a_n \end{bmatrix} \cdot \begin{bmatrix} e_q^{a_1} & 0 & 0 & 0 \\ 0 & e_q^{a_2} & 0 & 0 \\ 0 & 0 & \ddots & 0 \\ 0 & 0 & 0 & e_q^{a_n} \end{bmatrix} = \begin{bmatrix} b_1 & 0 & 0 & 0 \\ 0 & b_2 & 0 & 0 \\ 0 & 0 & \ddots & 0 \\ 0 & 0 & 0 & b_n \end{bmatrix}, \tag{96}$$

where $a_i$ are the eigenvalues of $A$ and $b_i$ are the eigenvalues of $B$. This implies the $k$-th eigenvalues of $A$ and $B$ are related by

$$a_k e_q^{a_k} = b_k \Rightarrow a_k = W_q(b_k). \tag{97}$$

Some matrices are simultaneously unitary and Hermitean, like the CNOT, Swap, Hadamard and Pauli quantum gates. Their eigenvalues are real with modulus equal to one: ±1. For CNOT gate, for example, one has

$$U_{CNOT} = \begin{bmatrix} 1 & 0 & 0 & 0 \\ 0 & 1 & 0 & 0 \\ 0 & 0 & 0 & 1 \\ 0 & 0 & 1 & 0 \end{bmatrix} = V \begin{bmatrix} -1 & 0 & 0 & 0 \\ 0 & 1 & 0 & 0 \\ 0 & 0 & 1 & 0 \\ 0 & 0 & 0 & 1 \end{bmatrix} V^\dagger \tag{98}$$

where $V$ is an unitary matrix whose columns are the eigenvectors of $U_{CNOT}$. Thus

$$U_{CNOT} e_q^{U_{CNOT}} = V \begin{bmatrix} -e_q^{-1} & 0 & 0 & 0 \\ 0 & e_q^1 & 0 & 0 \\ 0 & 0 & e_q^1 & 0 \\ 0 & 0 & 0 & e_q^1 \end{bmatrix} V^\dagger. \tag{99}$$

Since $U_{CNOT}\exp_q(U_{CNOT})$ in (99) is Hermitean, one can get the following family of two-qubit gates derived from CNOT.

$$U_q = e^{iU_{CNOT}e_q^{U_{CNOT}}} = V \begin{bmatrix} e^{-ie_q^{-1}} & 0 & 0 & 0 \\ 0 & e^{ie_q^1} & 0 & 0 \\ 0 & 0 & e^{ie_q^1} & 0 \\ 0 & 0 & 0 & e^{ie_q^1} \end{bmatrix} V^\dagger. \qquad (100)$$

Let us consider, now, that $B$ is a density matrix (Hermitean with $b_i \geq 0$ and $\Sigma_i b_i = 1$). In this case one has

$$A = V \begin{bmatrix} W_q(b_1) & 0 & 0 & 0 \\ 0 & W_q(b_2) & 0 & 0 \\ 0 & 0 & \ddots & 0 \\ 0 & 0 & 0 & W_q(b_n) \end{bmatrix} V^\dagger. \qquad (101)$$

Thus, the $q$-disentropy of $B$ is given by

$$D_q(B) = Tr(B^q A) = Tr\left[\left(Ae_q^A\right)^q A\right], \qquad (102)$$

where $Tr(X)$ is the trace of $X$.

## 9. Disentropy and Image Segmentation

One of the most basic tasks in image processing is the segmentation. It consists in to separate the main object of the background. In the simplest method this is done by establishing a threshold value, $t$. The segmented image is constructed setting all pixels with value smaller than $t$ to the value '0' (black) and all pixels with value larger or equal to $t$ to the value '255' (white). There are different techniques used to choose the best value of $t$ among them the maximal entropy method [29,32]. Let us assume, initially, an $N \times N$ image. Hence there are $N^2$ pixels. The value of the $k$-th pixel is $v_{(k)}$. Let us also assume that, for a given threshold value $t$, the set of pixels $A = \{a_1, a_2,\ldots,a_s\}$ is composed only by pixels with values larger or equal than $t$ while the set of pixels $B = \{b_1, b_2,…,b_r\}$ is composed only by pixels with values lower than $t$ ($r + s = N^2$). Then, the probability distributions for object $P(A)=\{p_a(1),…,p_a(k),…,p_a(s)\}$ and background $P(B)=\{p_b(1),…,p_b(k),…,p_b(r)\}$ can be constructed:

$$p_a(k) = v(a_k) \Big/ \sum_{i=1}^{s} v(a_i) \tag{103}$$

$$p_b(k) = v(b_k) \Big/ \sum_{i=1}^{r} v(b_i). \tag{104}$$

The $q$-entropy [8] of the distribution $P(A)$ ($P(B)$) is $S_q(A)$ ($S_q(B)$). The values of $S_q(A)$ and $S_q(B)$ depend on the value of $t$. The best value of $t$ is the one that maximizes $SAB_q = S_q(A) + S_q(B) + (1-q)S_q(A)S_q(B)$. The same method can be used employing the disentropy instead of entropy. The disentropy, eq. (11), of the distribution $P(A)$ ($P(B)$) is $D_q(A)$ ($D_q(B)$). The values of $D_q(A)$ and $D_q(B)$ depend on the value of $t$. The best value of $t$ is the one that minimizes $DAB_q = D_q(A) + D_q(B) - (1-q)D_q(A)D_q(B)$.

As an example, let us consider the Lena image. The disentropy and entropy versus threshold value $t$, for $q = 1/2$, can be seen in Fig. 10.

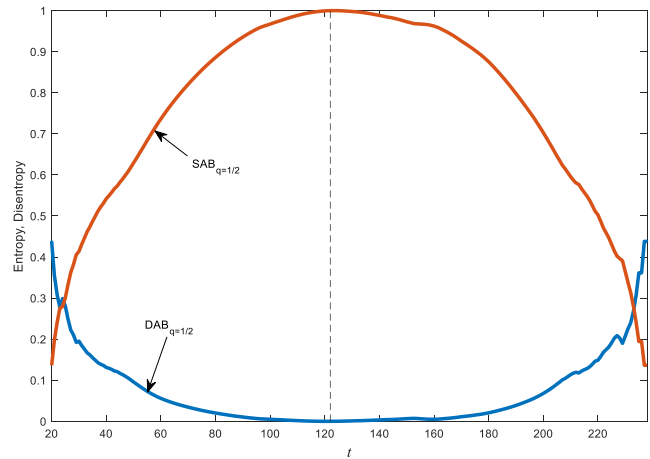

Fig. 10 – Normalized disentropy and $q$-entropy of the Lena image versus the threshold value $t$, for $q = 1/2$.

The original and segmented images using the threshold value $t = 121$, can be seen in Fig. 11.

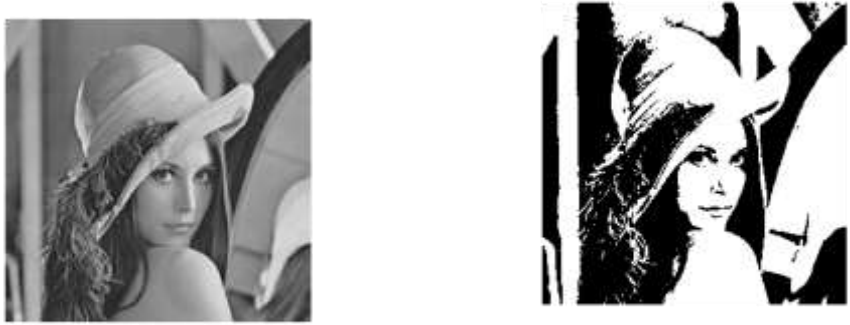

Fig. 11 – Original and segmented Lena images for $t = 121$ and $q = 1/2$.

Thus, the segmentation protocol can use the maximization of the entropy or the minimization of the disentropy.

## 10. Disentropy and integer numbers

The randomness is an important property found in nature and one can say it is crucial for life. For example, randomness is present in DNA assembly, games, cryptography and stock market, among others. A random process is characterized by its probability distribution. It is easy to determine if a probability distribution is maximally random or if it is deterministic (total absence of randomness). The uniform distribution is maximally random while the delta distribution has zero randomness. For both cases, one can use the disentropy ($D$) or the entropy ($S$) to identify maximally random and deterministic processes. The delta distribution has maximal disentropy ($D_q = 1$) and minimal entropy ($S_q = 0$) while the uniform distribution has maximal entropy ($S_q = 1$) and minimal disentropy ($D_q = 0$). So, the next question that arises is how much randomness exist in a process that is not deterministic neither maximally random, like the sequence generated by a biased coin. The coin tossing probability distribution for a biased coin is $P = \{p \rightarrow \text{head}, (1-p) \rightarrow \text{tail}\}$, with $p \neq 1/2$. If one uses only the entropy to measure the randomness, it is tempting to say that the distribution $P = \{p^*, (1-p^*)\}$ has "half maximally randomness" if $S(P^*) = ½$. However, in this case it is easy to note that $D(P^*) \neq ½$. The opposite is also true, if $P^+ = \{p^+, (1-p^+)\}$ is a distribution with "half maximally randomness" according to disentropy, $D(P^+) = ½$, then $S(P^+) \neq ½$.

Obviously, this happens because in general $D \neq 1-S$. In order to avoid this mismatch, one can use a measure of randomness based on disentropy and entropy. Using the normalized versions of disentropy and entropy one can define the randomness measure, $R_q$, named degree of randomness, given by

$$R_q = S_q - D_q. \tag{105}$$

Thus, $-1 \leq R_q \leq 1$. Equation (105) has an obvious meaning: $R_q < 0$ ($R_q > 0$) means the order (disorder) is dominant. A deterministic process has $R_q = -1$ while a maximally random process has $R_q = 1$. Figure 12 shows the degree of randomness ($R$), disentropy ($D$) and entropy ($S$) for the distribution $\{p, 1-p\}$ versus $p$. Figure 13 shows $R$, $D$ and $S$ for a Poissonian distribution ($p_n = e^{-\lambda}\lambda^n/n!$) versus its mean value $\lambda$. In both cases $q = 1$ was used. In Fig. 12, $R = 0$ for $p \sim 0.125125$ while in Fig. 13 $R = 0$ for $\lambda \sim 1.754$.

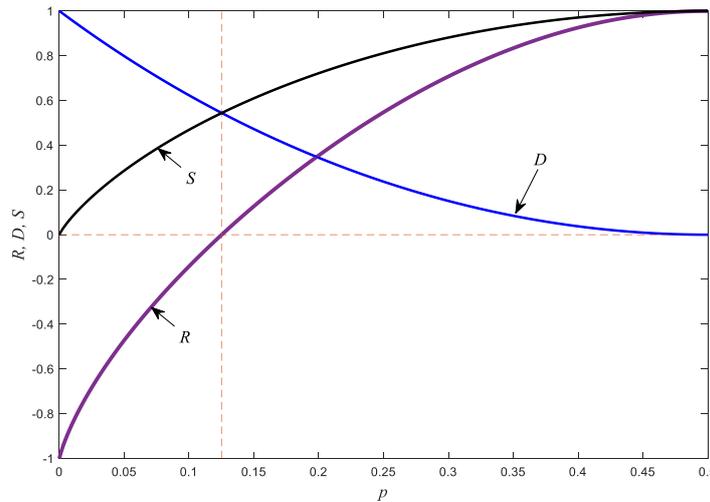

Fig. 12 – Degree of randomness ($R$), disentropy ($D$) and entropy ($S$) versus $p$ for the ditribution $\{p, (1-p)\}$ and $q = 1$.

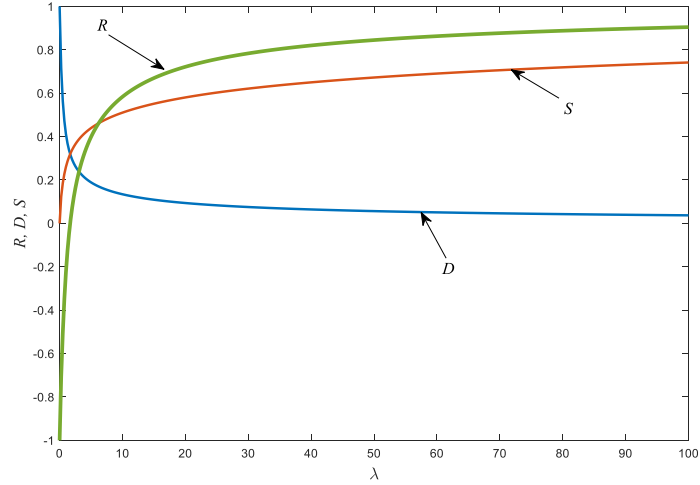

Fig. 13 – Degree of randomness (*R*), disentropy (*D*) and entropy (*S*) versus the mean value of a Poissonian distribution and $q = 1$.

Usually, statistical tests say, with an error probability lower than a threshold value, if a sequence of numbers is random or not, but they do not say how much randomness exist in that sequence. On the other hand, having a sequence of random numbers, one can estimate its probability density function (PDF) [33,34]. Having the PDF, the entropy and disentropy can be estimated and, hence, the degree of randomness of that sequence can also be estimated.

An interesting question that arises is: how much randomness does exist in the integer number *N*? In order to answer this question one has to define a probability distribution associated to number *N*. If the integer number *N* has the prime decomposition $N = p_1^{n_1} p_2^{n_2} \ldots p_k^{n_k}$, then the probability distribution associated to *N* is $P_N = \{log_N(p_1^{n_1}), log_N(p_2^{n_2}), \ldots, log_N(p_k^{n_k})\}$. Thus, the randomness associated to *N* is $R_q(N) = S_q(P_N) - D_q(P_N)$. Note that, if $N = p^k$ (prime power), then the distribution is a delta function and $R_q(p^k) = -1$. Moreover, one can create a quantum state associated a each integer number. The quantum states associated to prime powers are pure states:

$$\rho_p = U \begin{bmatrix} 0 & 0 & 0 & \cdots & 0 \\ 0 & \ddots & & & 0 \\ 0 & & 1 & & 0 \\ \vdots & & & \ddots & \vdots \\ 0 & 0 & 0 & \cdots & 0 \end{bmatrix} U^\dagger \qquad (106)$$

In (106) $U$ is a unitary matrix and the number 1 is placed in the element of the main diagonal associated to prime $p$: ($\rho_2(1,1) = 1$, $\rho_3(2,2) = 1$, $\rho_5(3,3) = 1$, etc...). Associated to $N = p_1^{n_1} p_2^{n_2} ... p_k^{n_k}$ is the mixed state

$$\rho_N = \sum_{k=1}^{K} \log_N\left(p_k^{n_k}\right) U \begin{bmatrix} 0 & 0 & 0 & \cdots & 0 \\ 0 & \ddots & & & 0 \\ 0 & & 1 & & 0 \\ \vdots & & & \ddots & \vdots \\ 0 & 0 & 0 & \cdots & 0 \end{bmatrix} U^{\dagger} = U \begin{bmatrix} 0 & 0 & 0 & \cdots & 0 \\ 0 & \log_N\left(p_1^{n_1}\right) & & & 0 \\ 0 & & \ddots & & 0 \\ \vdots & & & \log_N\left(p_K^{n_K}\right) & \vdots \\ 0 & 0 & 0 & \cdots & 0 \end{bmatrix} U^{\dagger} \quad (107)$$

As one may note in (107), the spectrum of $\rho_N$ is $\{log_N(p_1^{n_1}), log_N(p_2^{n_2}), ..., log_N(p_k^{n_k})\}$, hence, the degree of purity of $\rho_N$ (the difference between the entropy and disentropy of $\rho_N$) is equal to the randomness of $N$.

Since all prime powers have $R_q = -1$, one can use the degree of randomness as a distance measure between any integer number and the set of prime powers:

$$d_q(N, \mathscr{P}) = \frac{1 + R_q(N)}{2}. \quad (108)$$

In (108) $\mathscr{P}$ denotes the set of prime powers. Obviously, if $N = p^k$ then $d_q = 0$. In general, the distance between two integer number $N$ and $M$ can be defined by the relative disentropy between their distribution probabilities:

$$d_q(N, M) = D_q(P_N \| P_M). \quad (109)$$

If one uses (24.a), for example, then the distance between $N$ and the prime power set is

$$d_q(N, \mathscr{P}) = \min_j \left[ t_j^q |W_q(t_j) - W_q(1)| + \sum_{\substack{k \\ k \neq j}} t_k^q W_q(t_k) \right]. \quad (110)$$

In (110) $t_k$ are the probabilities associated to the integer $N$. Like eq. (108), the disentropic distance in (110) will be zero only for prime powers. From (108) and (110) one can see that integer numbers of the form $N = 2p$ (where $p$ is a prime much larger than 2) are the closest numbers from the prime power set, while number of the form $N = 2 \cdot 3 \cdot 5 \cdot 7 \cdot 11 \cdot 13 \cdot 17...$ (product of ordered sequence of prime numbers) are the farest numbers from the prime power set.

## 11. Perspectives and Conclusion

The disentropy has as many applications as entropy. It offers a different point of view, but it can also provide new results. In general, in problems where the entropy should be maximized (minimized), the disentropy should be minimized (maximized). The fact that the Lambert function and its generalizations can return a real value when its argument is negative (inside of an appropriate range) makes the disentropy very useful in the study of quasi-probability functions hence, the disentropy can bring new insights in the quantum-classical transition problem. The conditions required by $A$ and $B$ in order to $A\exp_q(A) = B$ to be satisfied, when $A$ and $B$ are operators or square matrices, are: I) $A$ and $B$ are Hermiteans. II) $[A,B] = 0$. III) $1/(1-q) \in \mathbb{Z}$. IV) The $n$-th eigenvalues of $A$ and $B$ are related by $a_n = W_q(b_n)$. Especial care must be taken with the allowed values of $q$ and the domains of the functions $exp_q$ and $W_q$. For example, $Ae_2^A = U_{CNOT}$ has no solution for $A$ since $W_2(-1)$ is not defined. Similarly, $\widehat{N}e_2^{\widehat{N}}$ cannot be calculated because $e_2^n$ is equal to zero for $n > 0$. At last, one should realize that it is hard to get analytical results with disentropy once the Lambert-Tsallis function, for example, does not have the sum-product relation found in the logarithmic function.

## Acknowledgements

This study was financed in part by the Coordenação de Aperfeiçoamento de Pessoal de Nível Superior - Brasil (CAPES) - Finance Code 001, and CNPq via Grant no. 307184/2018-8. Also, this work was performed as part of the Brazilian National Institute of Science and Technology for Quantum Information.